\documentclass[12pt]{article}

\usepackage{amsmath}
\usepackage{fullpage}
\usepackage{amssymb}
\usepackage{amsthm}
\usepackage{hyperref}
\usepackage{setspace}
\usepackage{graphicx}
\usepackage{bm}
\usepackage{physics}
\usepackage{authblk}

\begin{document}

 \title{The chemical birth-death process with additive noise}
 \author{John J. Vastola}
\affil{Department of Physics and Astronomy, Vanderbilt University, \\ Nashville, Tennessee}
 \maketitle

\begin{abstract}
The chemical birth-death process, whose chemical master equation (CME) is exactly solvable, is a paradigmatic toy problem often used to get intuition for how stochasticity affects chemical kinetics. In a certain limit, it can be approximated by an Ornstein-Uhlenbeck-like process which is also exactly solvable. In this paper, we use this system to showcase eight qualitatively different ways to exactly solve continuous stochastic systems: (i) integrating the stochastic differential equation; (ii) computing the characteristic function; (iii) eigenfunction expansion; (iv) using ladder operators; (v) the Martin-Siggia-Rose-Janssen-De Dominicis path integral; (vi) the Onsager-Machlup path integral; (vii) semiclassically approximating the Onsager-Machlup path integral; and (viii) approximating the solution to the corresponding CME. 
\end{abstract}

\section{Introduction}
\label{sec:intro}

What is there left to say about the Ornstein-Uhlenbeck process? First written down by Langevin \cite{langevin1908, lemons1997} and later studied in detail by Ornstein and Uhlenbeck \cite{ou1930}, it has become the prototypical toy problem for continuous stochastic dynamics, and is treated thoroughly in many textbooks \cite{risken1996, gardiner2009, vankampen2007}. It has been generalized to incorporate fractional diffusion \cite{lim2007, yan2008, eab2014}, time delay \cite{giuggioli2016}, and active behavior \cite{caprini2019}. Among other things, it has been used to model Brownian particles experiencing friction \cite{ou1930, gillespie1996}, Johnson noise \cite{gillespie1996}, harmonically trapped particles \cite{grebenkov2014}, heat baths \cite{pereira2015}, stock option prices \cite{perell2008}, pedestrian movement \cite{tordeux2016}, and active galactic nuclei \cite{kumar2015, takata2018}.

In this paper, we consider a related problem: the chemical\footnote{We will call this problem the \textit{chemical} birth-death process to distinguish it from the many other birth-death processes that have been considered in the literature, e.g. \cite{ismail1988, crawford2012, kononovicius2019}.} birth-death process with additive noise, which is defined by the stochastic differential equation (SDE)
\begin{equation} \label{eq:bdaddnoise}
\dot{x} = k - \gamma x + \sigma \eta(t)
\end{equation}
where $\eta(t)$ is a Gaussian white noise term, $x \in (-\infty, \infty)$, and $k, \gamma, \sigma > 0$. While Eq. \ref{eq:bdaddnoise} can trivially be changed into an Orstein-Uhlenbeck process by defining $y := x - k/\gamma$, we will focus on it as-is, because it has much to say about the relationship between different stochastic models, and because it is a good problem for illustrating the analytic tools in our arsenal for solving continuous stochastic problems. 

It is related to the chemical birth-death process, whose defining chemical reactions are
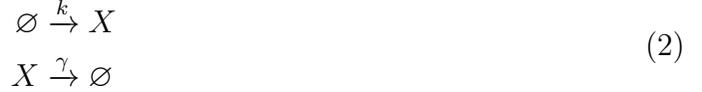
\begin{equation} \label{eq:rxns}
\begin{split}
\varnothing &\xrightarrow{k} X \\
X &\xrightarrow{\gamma} \varnothing 
\end{split}
\end{equation}
and whose corresponding chemical master equation (CME) reads
\begin{equation} \label{eq:CME}
\frac{\partial P(n, t)}{\partial t} = \ k \left[ P(n-1,t) - P(n,t) \right] + \gamma \left[ (n+1) P(n+1, t) - n P(n, t) \right] 
\end{equation}
where $P(n, t)$ is the probability that the system has $n$ $X$ molecules at time $t$ (with $n \in \{0, 1, 2, ...\}$). Although real biology is clearly more complicated, Eq. \ref{eq:CME} can be used as a first-pass model of how stochasticity influences mRNA or protein counts when gene-gene interactions are negligible \cite{fox2017, bressloff2017, munsky2018}. 

Specifically, Eq. \ref{eq:bdaddnoise} is related to Eq. \ref{eq:CME} by two approximations. The first approximation is to move to the continuous regime, and to approximate (\textit{a la} Gillespie \cite{gillespie2000, gillespie2013, schnoerr2017}) the dynamics of the CME via the chemical Langevin equation (CLE)
\begin{equation} \label{eq:CLE}
\dot{x} = k - \gamma x + \sqrt{k + \gamma x} \ \eta(t)
\end{equation}
where $x \in \left[ -\frac{k}{\gamma}, \infty \right)$ in this model, since the noise function has a nonzero probability of pushing the system into negative concentrations while its magnitude is nonzero. The second approximation is to suppose that we are sufficiently close to the steady state of the system, $\mu := \frac{k}{\gamma}$, so that $\epsilon := \frac{x - \mu}{\mu}$ is small. The noise function in Eq. \ref{eq:CLE} can then be approximated as
\begin{equation}
\begin{split}
\sqrt{k + \gamma x} = \sqrt{2k} \sqrt{1 + \frac{\epsilon}{2}} \approx \sqrt{2k} \left[ 1 + \frac{\epsilon}{4} \right] = \sqrt{2k} + \frac{\sqrt{2k}}{4\mu} (x - \mu) = \sigma_a + \sigma_m (x - \mu) \ .
\end{split}
\end{equation}
Keeping both terms corresponds to a (linear) multiplicative noise approximation, while only keeping the first term (i.e. assuming $x - \mu \approx 0$) corresponds to an additive noise approximation. Although we will keep the $\sigma$ from Eq. \ref{eq:bdaddnoise} arbitrary, the above argument shows that the choice of $\sigma$ that best approximates the dynamics of the CME (supposing the assumptions of the CLE hold, and that we are sufficiently close to the steady state $x_{ss} = \mu$) is
\begin{equation}
\sigma_{best} = \sqrt{2 k} \ .
\end{equation}
Hence, we recover Eq. \ref{eq:bdaddnoise}, and the domain expands to $x \in (-\infty, \infty)$, since a constant noise function has a nonzero probability of pushing a cell to arbitrarily negative concentrations. The transition probability corresponding to Eq. \ref{eq:bdaddnoise} is
\begin{equation} \label{eq:ptrans}
P(x, t; x_0, t_0) = \sqrt{\frac{\gamma}{\pi \sigma^2 (1 - e^{- 2 \gamma T})}} \exp\left[ - \frac{\gamma}{\sigma^2} \frac{[x - x_0 e^{- \gamma T} - \mu (1 - e^{- \gamma T})]^2}{(1 - e^{- 2 \gamma T})} \right]
\end{equation}
where $T := t - t_0$. Since all questions one might ask (e.g. moments and first-passage times) about the system described by Eq. \ref{eq:bdaddnoise} can be answered using $P(x, t; x_0, t_0)$, understanding the chemical birth-death process with additive noise in some sense reduces to computing and analyzing Eq. \ref{eq:ptrans}. 

For general stochastic systems, solving for $P(x, t; x_0, t_0)$ is usually a nontrivial task that requires employing all sorts of mathematical tools. In this paper, we will solve for Eq. \ref{eq:bdaddnoise} in eight qualitatively different ways---partially to showcase various methods, and partially to offer explicit examples of stochastic path integral \cite{vastola2019} computations, which seem to be rare in the literature. 

In Sec. \ref{sec:directsde} and \ref{sec:characteristics}, we describe two typical textbook approaches. In Sec. \ref{sec:eigexpansion} and \ref{sec:ladder}, we describe two approaches that mimic strategies usually used to solve the quantum harmonic oscillator. In Sec. \ref{sec:MSRJD} through \ref{sec:WKB}, we describe three path integral approaches, one of which (the Martin-Siggia-Rose-Janssen-De Dominicis path integral, used in Sec. \ref{sec:MSRJD}) is particularly straightforward. Finally, in Sec. \ref{sec:CMEcomp} we derive Eq. \ref{eq:bdaddnoise} by approximating the solution to the CME (Eq. \ref{eq:CME}) in the large $\mu$ limit. 

\section{Direct SDE solution method}
\label{sec:directsde}
The additive noise birth-death process is simple enough that we can solve its SDE directly, in exactly the same way the Ornstein-Uhlenbeck SDE is usually solved. Because the solution has a special form (i.e. a normal distribution), we can then use that solution to find the transition probability. This approach is textbook material; see Gardiner \cite{gardiner2009}  for a reference. 

In mathematicians' notation, our SDE reads
\begin{equation}
dx = (k - \gamma x) dt + \sigma dW 
\end{equation}
where $W$ is a Wiener process. The trick is to use the `integrating factor' $e^{\gamma t}$ to eliminate the $x$-dependence from the right-hand side of the SDE:
\begin{equation}
\begin{split}
d(e^{\gamma t} x) = \gamma e^{\gamma t} x \ dt + e^{\gamma t} \ dx = \gamma e^{\gamma t} x \ dt + e^{\gamma t} \left[ (k - \gamma x) \ dt + \sigma \ dW \right] = k e^{\gamma t} \ dt + \sigma e^{\gamma t} \ dW \ .
\end{split}
\end{equation}
Integrating both sides, we have
\begin{equation}
\begin{split}
e^{\gamma t} x(t) &= x_0 e^{\gamma t_0} + k \int_{t_0}^t e^{\gamma t'}  \ dt' + \sigma \int_{t_0}^t e^{\gamma t'} \ dW_{t'} \ .
\end{split}
\end{equation}
By the definition of the Ito integral \cite{ito1944, gardiner2009}, we have
\begin{equation} \label{eq:Itointegral}
\begin{split}
\int_{t_0}^t e^{\gamma t'} \ dW_{t'}  &:= \lim_{N \to \infty} \sum_{j=1}^N e^{\gamma [t_0 + (j-1) \Delta t]}  r_j(0, \Delta t)
\end{split}
\end{equation}
where the $r_j$ are normal random variables with mean $0$ and variance $\Delta t := (t - t_0)/N$. Using the usual rules for manipulating linear combinations of normal random variables \cite{gillespie2000} (and reusing the $r_j$ labels for convenience), we have
\begin{equation}
\sum_{j=1}^N e^{\gamma [t_0 + (j-1) \Delta t]}  r_j(0, \Delta t) = \sum_{j=1}^N  r_j\left(0, e^{2 \gamma [t_0 + (j-1) \Delta t]} \Delta t \right) =   R\left(0, \sum_{j=1}^N e^{2\gamma[t_0 + (j-1) \Delta t]} \Delta t \right) 
\end{equation}
where $R$ is a normal random variable with mean $0$ and variance
\begin{equation}
\sum_{j=1}^N e^{2\gamma [t_0 + (j-1) \Delta t]} \Delta t \approx \int_{t_0}^t e^{2\gamma t'} \ dt' = \frac{e^{2\gamma t} - e^{2\gamma t_0}}{2\gamma} 
\end{equation}
which becomes exact in the $N \to \infty$ limit. The other integral is just 
\begin{equation}
\int_{t_0}^t e^{\gamma t'}  \ dt' = \frac{e^{\gamma t} - e^{\gamma t_0}}{\gamma}
\end{equation}
so we have 
\begin{equation}
\begin{split}
x(t) &= x_0 e^{- \gamma T} + \mu \left[ 1 - e^{-\gamma T}  \right] + \sigma e^{-\gamma t} R\left( 0, \frac{e^{2\gamma t} - e^{2\gamma t_0}}{2\gamma}  \right)
\end{split}
\end{equation}
where $T := t - t_0$. Again using what we know about linear combinations of normal random variables, this can be rewritten as
\begin{equation}
\begin{split}
x(t) &= R\left( x_0 e^{- \gamma T} + \mu \left[ 1 - e^{-\gamma T}  \right] , \frac{\sigma^2}{2\gamma} \left[ 1 - e^{-2\gamma T} \right] \right) 
\end{split}
\end{equation}
where $R$ is normally distributed with the above mean and variance. Since
\begin{equation}
x(t) \sim \mathcal{N}\left( x_0 e^{- \gamma T} + \mu \left[ 1 - e^{-\gamma T}  \right] , \frac{\sigma^2}{2\gamma} \left[ 1 - e^{-2\gamma T} \right] \right)
\end{equation}
we recover Eq. \ref{eq:ptrans} by writing down a normal distribution with the above mean and variance. If we are just interested in moments, we do not even have to calculate the Ito integral (Eq. \ref{eq:Itointegral}) \cite{gardiner2009}; instead, we can use
\begin{equation}
\begin{split}
x(t) &= x_0 e^{- \gamma T} + \mu \left[ 1 - e^{-\gamma T}  \right] + \sigma e^{-\gamma t} \int_{t_0}^t e^{\gamma t'} \ dW_{t'} 
\end{split}
\end{equation}
and the properties of Ito integrals/white noise. For example,
\begin{equation}
\begin{split}
\expval{x(t)} &= \expval{x_0 e^{- \gamma T} + \mu \left[ 1 - e^{-\gamma T}  \right]} + \sigma e^{-\gamma t} \expval{\int_{t_0}^t e^{\gamma t'} \ dW_{t'} } = x_0 e^{- \gamma T} + \mu \left[ 1 - e^{-\gamma T}  \right] 
\end{split}
\end{equation}
and
\begin{equation}
\begin{split}
\mathrm{Var}[x(t)] &= \sigma^2 e^{-2 \gamma t} \expval{ \int_{t_0}^t e^{\gamma t'} \ dW_{t'} \int_{t_0}^t e^{\gamma s'} \ dW_{s'} } = \sigma^2 e^{-2 \gamma t}  \int_{t_0}^t e^{2\gamma t'} \ dt' = \frac{\sigma^2}{2 \gamma} \left[ 1  -  e^{-2 \gamma T}  \right]  \ .
\end{split}
\end{equation}

\section{Method of characteristics}
\label{sec:characteristics}

The time-dependent probability density $P(x, t)$ of the stochastic system described by Eq. \ref{eq:bdaddnoise} satisfies the Fokker-Planck equation, which reads
\begin{equation} \label{eq:TDFP}
\frac{\partial P(x,t)}{\partial t} = - \frac{\partial}{\partial x} \left[ (k - \gamma x) P(x,t) \right] + \frac{\sigma^2}{2} \frac{\partial^2 P(x,t)}{\partial x^2} \ .
\end{equation}
The boundary conditions are:
\begin{enumerate}
\item $\lim_{x \to \pm \infty} P(x, t) = 0$, $\lim_{x \to \pm \infty} P'(x, t) = 0$,  and $P(x, t)$ dies off fast enough that the integral $\int P(x, t) dx$ converges for all $t$.
\item $P(x,0) = P_0(x)$ for some initial distribution $P_0(x)$.
\end{enumerate} 
We need no normalization requirement, since if $P_0(x)$ is normalized, the first condition guarantees that $P(x, t)$ will remain normalized for all times $t$. Because we would like to calculate the transition probability $P(x, t; x_0, t_0)$, we will be interested in the initial condition $P_0(x) = \delta(x - x_0)$. 

There are many ways to solve the Fokker-Planck equation; in this section, we will consider taking its Fourier transform. This corresponds to computing the characteristic function of our system 
\begin{equation}
G(q, t) := \expval{e^{i q x}} =  \int_{-\infty}^{\infty} dq \ e^{i q x} P(x, t) 
\end{equation}
which is equivalent to the original probability density function (since one can be recovered from the other by a Fourier/inverse Fourier transform). Taking the Fourier transform of both sides of Eq. \ref{eq:TDFP}, we find that $G(q, t)$ satisfies
\begin{equation} \label{eq:GDE}
\frac{\partial G(q, t)}{\partial t} + \gamma q \frac{\partial G(q, t)}{\partial q} = \left[ i k q - \frac{\sigma^2}{2} q^2 \right] G(q, t)
\end{equation}
subject to the initial condition\footnote{The decay at infinity conditions are already taken into account by assuming that it is possible to Fourier transform $P(x, t)$.}
\begin{equation} \label{eq:GIC}
G(q, t_0) = \int_{-\infty}^{\infty} dq \ e^{i q x} \delta(x - x_0) = e^{i q x_0} \ .
\end{equation}
The method of characteristics \cite{risken1996, courant2008}, offers a way to solve first-order partial differential equations (PDEs) like this one. It involves supposing that there are parameterized curves along with the PDE reduces to an ordinary differential equation (ODE). In particular, we suppose that $q = q(s)$, $t = t(s)$, and
\begin{equation} \label{eq:totalderiv}
\begin{split}
\frac{d}{ds} [ G(q(s),t(s)) ] &= \frac{\partial G(q, t)}{\partial t} \frac{dt}{ds} + \frac{\partial G(q, t)}{\partial q} \frac{dq}{ds} = \frac{\partial G(q, t)}{\partial t} + \gamma q \frac{\partial G(q, t)}{\partial q}
\end{split}
\end{equation}
i.e. that
\begin{equation}
\begin{split}
\frac{dt}{ds} = 1 \ , \ \frac{dq}{ds} = \gamma q  \ .
\end{split}
\end{equation}
The solutions of these ODEs are
\begin{equation} \label{eq:tandq}
\begin{split}
t(s) &= t_0 + s \\
q(s) &= q_0 e^{\gamma s}
\end{split}
\end{equation}
where we have chosen our parameterization so that $s = 0$ corresponds to $t_0$, and $s = T = t - t_0$ corresponds to $t(s) = t$. 

Substituting Eq. \ref{eq:totalderiv} and Eq. \ref{eq:tandq} into Eq. \ref{eq:GDE} yields the first-order ODE
\begin{equation}
\frac{d G(s)}{ds} = \left[ i k q_0 e^{\gamma s} - \frac{\sigma^2}{2} q_0^2 e^{2\gamma s} \right] G(s) \ .
\end{equation}
Proceed by separation of variables to get
\begin{equation}
\begin{split}
\int_0^T \frac{dG}{G} = \int_0^T i k q_0 e^{\gamma s} - \frac{\sigma^2}{2} q_0^2 e^{2\gamma s} \ ds 
\end{split}
\end{equation}
which is easily solved to obtain
\begin{equation}
\begin{split}
G(T) &= G_0 \exp\left\{  i \mu q_0 \left[ e^{\gamma T} - 1 \right]- \frac{\sigma^2}{4 \gamma} q_0^2 \left[ e^{2\gamma T} - 1\right] \right\} \\
&= G_0 \exp\left\{  i \mu q \left[ 1 - e^{-\gamma T}  \right]- \frac{\sigma^2}{4 \gamma} q^2 \left[ 1 - e^{- 2\gamma T} \right] \right\} 
\end{split}
\end{equation}
where we have used that $q_0 = q(T) e^{- \gamma T}$. At this point, all we have to do is incorporate our initial condition (Eq. \ref{eq:GIC}). Doing so, we have
\begin{equation}
G(s=0) = G_0 = e^{i q(0) x_0} = e^{i q x_0 e^{- \gamma T}}
\end{equation}
so that our final answer is
\begin{equation} \label{eq:char_answer}
G(q, t) = \exp\left\{ i q \left[ x_0 e^{- \gamma T} + \mu \left( 1 - e^{- \gamma T} \right) \right] - \frac{1}{2} q^2 \left[ \frac{\sigma^2}{2 \gamma} \left( 1 - e^{- 2 \gamma T} \right) \right]  \right\} \ .
\end{equation}
This may look familiar; the characteristic function of a normal distribution with mean $\bar{\mu}$ and variance $\bar{\sigma}^2$ is
\begin{equation}
\exp\left[ i \bar{\mu} q - \frac{\bar{\sigma}^2}{2} q^2  \right] \ .
\end{equation}
Comparing this with Eq. \ref{eq:char_answer}, we find that $P(x, t; x_0, t_0)$ must be a normal distribution with mean and variance given by
\begin{equation}
\begin{split}
\bar{\mu} &:= x_0 e^{- \gamma T} + \mu \left( 1 - e^{- \gamma T} \right) \\
\bar{\sigma}^2 &:= \frac{\sigma^2}{2 \gamma} \left( 1 - e^{- 2 \gamma T} \right)
\end{split}
\end{equation}
i.e. we recover Eq. \ref{eq:ptrans}. Alternatively, we can just inverse Fourier transform Eq. \ref{eq:char_answer} to recover Eq. \ref{eq:ptrans}.

\section{Eigenfunction expansion method}
\label{sec:eigexpansion}

Applying the standard separation of variables ansatz $P(x, t) = P_E(x) T(t)$ to the Fokker-Planck equation (Eq. \ref{eq:TDFP}) yields the general solution
\begin{equation} \label{eq:FPgenslnE}
P(x, t) = \sum_{E} c_E P_E(x) e^{- E (t - t_0)}
\end{equation}
where the $c_E$ are chosen so that $P(x, t_0)$ equals some initial distribution $P_0(x)$, and where $P_E(x)$ satisfies the \textit{time-independent} Fokker-Planck equation
\begin{equation} \label{eq:TIFP}
-E P_E(x) = \gamma P_E(x) - (k - \gamma x) \frac{\partial P_E(x)}{\partial x}  + \frac{\sigma^2}{2} \frac{\partial^2 P_E(x)}{\partial x^2} 
\end{equation}
for some constant $E \geq 0$. This eigenfunction expansion technique is discussed by Risken \cite{risken1996} in his monograph on the Fokker-Planck equation, and is occasionally used in the literature \cite{cukier1973, vanderMee1987, hu1990, zhang1997, koide2017}.

\subsection{Steady state Fokker-Planck solution}

As a starting point, we would like to find $P_{ss}(x)$, the steady state solution to the Fokker-Planck equation. Setting $\partial P/\partial t = 0$ in Eq. \ref{eq:TDFP}, we have
\begin{equation} \label{eq:ssFPd}
0 = - \frac{\partial}{\partial x} \left[ (k - \gamma x) P_{ss}(x) \right] + \frac{\sigma^2}{2} \frac{\partial^2 P_{ss}(x)}{\partial x^2} \ .
\end{equation}
Integrate both sides (and note that the arbitrary constant that appears must be zero for both sides to vanish at infinity) to obtain the steady state Fokker-Planck equation
\begin{equation} \label{eq:ssFP}
0 = (k - \gamma x) P_{ss}(x)  - \frac{\sigma^2}{2} P'_{ss}(x) \ .
\end{equation}
Solving this simple ODE and normalizing our result, we obtain
\begin{equation} \label{eq:pss}
P_{ss}(x) = \sqrt{ \frac{\gamma}{\pi \sigma^2} } \exp\left[ - \frac{\gamma}{\sigma^2} \left( x - \mu \right)^2  \right] \ .
\end{equation}

\subsection{Eigenfunctions}

Assume that the solution to Eq. \ref{eq:TIFP} can be written $P_E(x) = Q_E(x) P_{ss}(x)$. Substituting this ansatz into Eq. \ref{eq:TIFP} and using Eq. \ref{eq:ssFP} to simplify the result yields the equation
\begin{equation}
0 = \frac{\sigma^2}{2}  Q_E''(x) + (k - \gamma x) Q_E'(x)  + E Q_E(x)   \ .
\end{equation}
for $Q_E(x)$. Changing variables to $y := \sqrt{\gamma/\sigma^2} (x - \mu)$, our equation becomes
\begin{equation} \label{eq:qDE}
0 = Q_E''(y) - 2 y Q_E'(y) + 2 \bar{E} Q_E(y) 
\end{equation}
where $\bar{E} := E/\gamma$. A standard power series analysis\footnote{This sort of analysis is often done in textbook studies of the quantum harmonic oscillator, so we will not reproduce it here. See Griffiths \cite{griffiths2018} for one example.} of Eq. \ref{eq:qDE} shows that it will only have solutions which do not blow up at infinity provided that $\bar{E}$ is a nonnegative integer $n$. Hence, Eq. \ref{eq:qDE} is just Hermite's differential equation, so its solutions can be written
\begin{equation} \label{eq:qsln}
Q_n(x) = \sqrt{ \frac{1}{2^n n!} } H_n\left( \sqrt{\gamma/\sigma^2} (x - \mu) \right) 
\end{equation}
where the prefactor is chosen for our later convenience. While the $P_n = Q_n P_{ss}$ functions can obviously not be interpreted as probability distributions in their own right, since $Q_n$ sometimes takes negative values, they do convey information about the relative probability of different transient solutions (i.e. solutions whose time-dependence goes like $e^{- \gamma n T}$).

\subsection{The propagator}
\label{sec:propagatorsum}

Using Eq. \ref{eq:FPgensln} and Eq. \ref{eq:qsln}, the general solution to the Fokker-Planck equation is
\begin{equation} \label{eq:FPgensln}
P(x, t) = \sum_{n=0}^\infty c_n P_n(x) e^{- E_n (t - t_0)} = \sum_{n=0}^\infty c_n Q_n(x) P_{ss}(x) e^{- \gamma n T}
\end{equation}
with the constants $c_n$ chosen to match the assumed initial distribution $P_0(x)$. To actually calculate the $c_n$, one can invoke the orthogonality of the Hermite polynomials, which reads 
\begin{equation}
\int_{-\infty}^{\infty} H_m(y) H_n(y) e^{- y^2} \ dy = \sqrt{\pi} 2^n n! \ \delta_{nm} \ ,
\end{equation}
to say that
\begin{equation} \label{eq:qORTHO}
\int_{-\infty}^{\infty} Q_m(x) Q_n(x) P_{ss}(x) \ dx = \delta_{nm}
\end{equation}
i.e. that the $Q_n$ are orthonormal with respect to the weight function $P_{ss}(x)$. Let's exploit this relationship to compute the coefficients $c_n$ for the initial distribution $P_0(x) = \delta(x - x_0)$. At $t = t_0$, we have
\begin{equation}
\delta(x - x_0) = \sum_{n=0}^\infty c_n Q_n(x) P_{ss}(x) \ .
\end{equation}
Multiply both sides by $Q_m(x)$ and integrate, using Eq. \ref{eq:qORTHO}. We get $c_m = Q_m(x_0)$, so our solution is
\begin{equation} \label{eq:Hexpansion}
P(x, t; x_0, t_0) = \sqrt{ \frac{\gamma}{\pi \sigma^2} } \sum_{n=0}^\infty \frac{1}{2^n n!} H_n\left( \sqrt{ \frac{\gamma}{\sigma^2}} (x_0 - \mu) \right) H_n\left( \sqrt{ \frac{\gamma}{\sigma^2}} (x - \mu) \right)   e^{ - \frac{\gamma}{\sigma^2} \left( x - \mu \right)^2 } e^{- \gamma n T} \ .
\end{equation}
To sum this, we can either use the integral representation \cite[Eq.~18.10.10]{DLMF}
\begin{equation}
H_n(y) = \frac{(-2 i)^n e^{y^2}}{\sqrt{\pi}} \int_{-\infty}^{\infty} e^{- t^2} t^n e^{2 i y t} \ dt
\end{equation}
of the Hermite polynomials, or we can explicitly invoke Mehler's formula \cite{mehler1866, bateman1953}, which says that
\begin{equation}
\sum_{n=0}^\infty \frac{(\rho/2)^n}{n!} ~ \mathit{H}_n(y_0)\mathit{H}_n(y) =\frac 1{\sqrt{1-\rho^2}}\exp\left(-\frac{\rho^2 (y_0^2+y^2)- 2\rho \ y_0 y}{1-\rho^2}\right) \ .
\end{equation}
Either way, we recover Eq. \ref{eq:ptrans} for $P(x, t; x_0, t_0)$. 

\section{Ladder operator method}
\label{sec:ladder}

Ladder (or \textit{raising and lowering}, or \textit{creation and annihilation}) operators  facilitate a straightforward treatment of the quantum harmonic oscillator in elementary quantum mechanics \cite{griffiths2018}. Analogous methods have been used to solve the Fokker-Planck equation \cite{titulaer1978, bernstein1984, arvedson2006, koide2017}, although the approach seems to work best for simple systems (e.g. one-dimensional or having a linear drift term). 

\subsection{Basic formalism}

As in the previous section, we begin with a separation of variables ansatz, and seek to solve the time-independent Fokker-Planck equation (Eq. \ref{eq:TIFP}). For this approach, we will (as in \cite{vastola2019}) work in a Hilbert space consisting of states
\begin{equation}
\ket{\phi} = \int_{-\infty}^{\infty} dx \ c(x) \ket{x}
\end{equation}
and introduce operators $\hat{x}, \hat{p}$ and $\hat{H}$ that act as
\begin{equation}
\begin{split}
\hat{x} \ket{\phi} &:= \int_{-\infty}^{\infty} dx \ x c(x) \ket{x} \ , \ \hat{p} \ket{\phi} := \int_{-\infty}^{\infty} dx \ - \frac{\partial c(x)}{\partial x} \ket{x} \ , \ \hat{H}  := \hat{p} f(\hat{x}) + \frac{\sigma^2}{2} \hat{p}^2 
\end{split}
\end{equation}
on a general state $\ket{\phi}$.  Motivated by the quantum harmonic oscillator, we will also introduce the creation and annihilation operators
\begin{equation}
\hat{a}^+ := \hat{p} \ , \ \hat{a} :=  \hat{x} - \mu - \frac{\sigma^2}{2 \gamma} \hat{p} \ .
\end{equation}
We stress that the operators $a$ and $a^+$ are \textit{not} Hermitian conjugates of each other; their conjugates will be determined in the next subsection. Since $[\hat{x}, \hat{p}] = 1$, $\hat{a}$ and $\hat{a}^+$ satisfy the usual commutation relation
\begin{equation}
\begin{split}
[ \hat{a}, \hat{a}^+] = \left[\hat{x} - \mu - \frac{\sigma^2}{2 \gamma} \hat{p}, \ \hat{p} \right] = [\hat{x}, \hat{p}] = 1 \ .
\end{split}
\end{equation}
Also note that we can write the Hamiltonian as $\hat{H} = -\gamma \ \hat{a}^+ \hat{a}$, and that the time-independent Fokker-Planck equation says
\begin{equation}
\ket{\psi_E} := \int_{-\infty}^{\infty} dx \ P_E(x) \ket{x} \ \implies \ \hat{H} \ket{\psi_E} = - E \ket{\psi_E}
\end{equation}
i.e. $\ket{\psi_E}$ is an eigenstate of the Hamiltonian with eigenvalue $-E$. 

\subsection{Allowed energies}

Let's run through the usual ladder operator arguments. Suppose that $\ket{\psi_E}$ is an eigenstate of $\hat{H}$ with eigenvalue $-E$. First, note that acting on $\ket{\psi_E}$ with the annihilation operator $\hat{a}$ yields an eigenstate with eigenvalue $-E + \gamma$:
\begin{equation}
\hat{H} \ \hat{a} \ket{\psi_E} = - \gamma \hat{a}^+ \hat{a}  \hat{a} \ \ket{\psi_E}   = - \gamma (\hat{a}  \hat{a}^+ - 1 ) \hat{a}  \ \ket{\psi_E} = \hat{a} (\hat{H} + \gamma ) \ \ket{\psi_E}   = (-E + \gamma ) \ \hat{a} \ket{\psi_E}   \ .
\end{equation}
In just the same way, one can show that acting on $\ket{\psi_E}$ with the creation operator $\hat{a}^+$ yields an eigenstate with eigenvalue $-E - \gamma$. This means we can take any eigenstate and use it to generate new eigenstates with higher or lower energies $E$. Recall that our `biologically permissible' solutions need $E > 0$, or else they will blow up in time. To prevent our ladder operators for permitting such solutions, we need it to be the case that 
\begin{equation} \label{eq:gstate_condition}
\hat{a} \ket{\psi_E} = 0 
\end{equation}
for some eigenstate $\ket{\psi_E}$. But this is \textit{precisely} what is true for the steady state solution we found earlier---it is a solution with $E = 0$! Acting many times on this `ground state' with the creation operator yields states with energies $E_1 = \gamma$, $E_2 = 2 \gamma$, and so on. We obtain a countably infinite number of eigenstates with energies
\begin{equation}
E_n = n \gamma
\end{equation}
for $n = 0, 1, 2, ...$. This must be \textit{all} possible eigenstates, in fact: if there was an eigenstate whose energy was not an integer multiple of $\gamma$, we could use the annihilation operator $\hat{a}$ to construct an eigenstate with negative energy, which is not allowed.

\subsection{Proving orthonormality}

Label the allowed eigenstates $\ket{\psi_n}$, since we know now that there are only countably many of them. We have that
\begin{equation}
\ket{\psi_n} := C_n (a^+)^n \ket{\psi_0}
\end{equation}
where $\ket{\psi_0}$ is defined by Eq. \ref{eq:gstate_condition} and the constants $C_n$ are to be determined. We would like to be able to invoke the orthogonality of the $\ket{\psi_n}$ in order to construct a general solution to the time-dependent Fokker-Planck equation, Eq. \ref{eq:TDFP}. To do this, we will need to define an inner product and show that the $\ket{\psi_n}$ are orthogonal to each other with respect to it (and choose $C_n$ so that they are also normalized). 

Define an inner product by
\begin{equation}
\braket{x}{y} := \frac{\delta(x - y)}{P_{ss}(x)} 
\end{equation}
so that the inner product of two arbitrary states $\ket{\phi_1}$ and $\ket{\phi_2}$ reads
\begin{equation}
\begin{split}
\braket{\phi_1}{\phi_2} &= \int_{-\infty}^{\infty} dx \int_{-\infty}^{\infty} dy \ c_1^*(x) c_2(y) \braket{x}{y} = \int_{-\infty}^{\infty} dx  \ \frac{c_1^*(x) c_2(x)}{P_{ss}(x)} \ .
\end{split}
\end{equation}
Note that, with respect to this inner product, our `ground state' $\ket{\psi_0}$ is normalized:
\begin{equation}
\begin{split}
\braket{\psi_0}{\psi_0} &= \int_{-\infty}^{\infty} dx  \ \frac{P_{ss}(x) P_{ss}(x)}{P_{ss}(x)}  = 1 \ .
\end{split}
\end{equation}
It is not true that the creation and annihilation operators we defined, $\hat{a}$ and $\hat{a}^+$, are Hermitian conjugates with respect to this inner product. But they are \textit{almost} Hermitian conjugates, in the following sense. Note, for arbitrary states $\ket{\phi_1}$ and $\ket{\phi_2}$,
\begin{equation}
\begin{split}
\matrixel{\phi_1}{\hat{p}}{\phi_2} &= \int_{-\infty}^{\infty} dx  \ \frac{c_1^*(x)}{P_{ss}(x)} \left[ - \frac{dc_2}{dx} \right] = \int_{-\infty}^{\infty} dx  \ c_2(x) \frac{d}{dx} \left[ \frac{c_1^*(x)}{P_{ss}(x)} \right] 
\end{split}
\end{equation}
where we have integrated by parts and thrown away the boundary terms. The boundary terms do vanish for the states we care about, which have $c(x) \sim [\text{polynomial}] \cdot P_{ss}(x)$. Next, compute
\begin{equation}
\begin{split}
\frac{d}{dx} \left[ \frac{c_1^*(x)}{P_{ss}(x)} \right] &= \frac{\frac{dc_1^*}{dx}}{P_{ss}(x)} - \frac{P_{ss}'(x) c_1^*(x)}{[P_{ss}(x)]^2} = \frac{\frac{dc_1^*}{dx} + \frac{2 \gamma}{\sigma^2} \left( x - \mu  \right) c_1^*(x) }{P_{ss}(x)} \\
\end{split}
\end{equation}
where we have used Eq. \ref{eq:ssFP}. Then
\begin{equation}
\begin{split}
\matrixel{\phi_1}{\hat{p}}{\phi_2} &= \int_{-\infty}^{\infty} dx  \  \frac{c_2(x) \left[ \frac{dc_1^*}{dx} + \frac{2 \gamma}{\sigma^2} \left( x - \mu  \right) c_1^*(x)  \right]}{P_{ss}(x)} = \mel**{\phi_2}{- \hat{p} + \frac{2 \gamma}{\sigma^2} \left( \hat{x} - \mu  \right)}{\phi_1}^* \ .
\end{split}
\end{equation}
In other words,
\begin{equation} \label{eq:pconjugate}
(\hat{p})^{\dag} = - \hat{p} + \frac{2 \gamma}{\sigma^2} \left( \hat{x} - \mu  \right)
\end{equation}
with respect to our inner product. Using Eq. \ref{eq:pconjugate} and that $\hat{x}$ is Hermitian, we can show that
\begin{equation}
\begin{split}
(\hat{a}^+)^{\dag} = \frac{2 \gamma}{\sigma^2} \ \hat{a} \ , \ (\hat{a})^{\dag} = \frac{\sigma^2}{2 \gamma}  \hat{a}^+ \ .
\end{split}
\end{equation}
These results together mean that $\hat{H} := - \gamma \hat{a}^+ \hat{a}$ (along with the `number operator' $\hat{N} := \hat{a}^+ \hat{a}$) is Hermitian. This can be used to show (in a slick way) that the $\ket{\psi_n}$ are orthogonal:
\begin{equation}
\begin{split}
\braket{\psi_m}{\psi_n} =  \frac{\matrixel{\psi_m}{n}{\psi_n}}{n} = \frac{\matrixel{\psi_m}{\hat{N}}{\psi_n}}{n} = \frac{\matrixel{\psi_n}{\hat{N}}{\psi_m}^*}{n} = \frac{\matrixel{\psi_n}{m}{\psi_m}^*}{n} = \frac{m}{n} \braket{\psi_m}{\psi_n} 
\end{split}
\end{equation}
which for $m \neq n$ forces $\braket{\psi_m}{\psi_n} = 0$. Now we should normalize the $\ket{\psi_n}$. Note,
\begin{equation}
\begin{split}
\braket{\psi_n}{\psi_n} &= \left[ C_n (\hat{a}^+)^n \ket{\psi_n} \right]^{\dag} C_n (\hat{a}^+)^n \ket{\psi_n} = |C_n|^2 \left( \frac{2 \gamma}{\sigma^2} \right)^n \matrixel{\psi_0}{\hat{a}^n (\hat{a}^+)^n}{\psi_0} \ .
\end{split}
\end{equation}
Repeatedly use the facts that $[\hat{a}, (\hat{a}^+)^{j}] = j (\hat{a}^+)^{j-1}$ for $j \in \mathbb{N}$ and $\hat{a} \ket{\psi_0} = 0$ to obtain
\begin{equation}
\begin{split}
\braket{\psi_n}{\psi_n} &= |C_n|^2 \left( \frac{2 \gamma}{\sigma^2} \right)^n n! \braket{\psi_0}{\psi_0} = |C_n|^2 \left( \frac{2 \gamma}{\sigma^2} \right)^n n! 
\end{split}
\end{equation}
which means that we should choose $C_n = \frac{1}{\sqrt{n!}} \left( \frac{\sigma^2}{2 \gamma} \right)^{n/2}$. One can invoke the Rodrigues formula of the Hermite polynomials to show that the $\ket{\psi_n}$  match what we found earlier (c.f. Eq. \ref{eq:qsln}). 

\subsection{Final comments}

We can write
\begin{equation} \label{eq:ketsln}
\ket{\psi(t)} = \sum_{n = 0}^{\infty} c_n \ket{\psi_n} e^{- \gamma n T} \ ,
\end{equation}
exploit the orthonormality of the $\ket{\psi_n}$ to derive the $c_n$ corresponding to the transition probability, and sum the propagator as in Sec. \ref{sec:propagatorsum}. 

One last note about the ladder operator approach: just as in quantum mechanics, the creation and annihilation operators are useful for calculating moments. For example, since $\hat{x}$ can be written as 
\begin{equation} \label{eq:xviaa}
\hat{x} = \frac{\sigma^2}{2 \gamma} \hat{a}^+ + \hat{a} + \mu 
\end{equation}
we can compute
\begin{equation}
\expval{x} = \matrixel{ \psi_0 }{\hat{x}}{\psi(t)} = \int_{-\infty}^{\infty} dx \ x P(x, t)
\end{equation}
by using Eq. \ref{eq:ketsln} and the properties
\begin{equation}
\begin{split}
\hat{a}^+ \ket{\psi_n} = \sqrt{\frac{2 \gamma}{\sigma^2}} \sqrt{n+1} \ket{\psi_{n+1}} \ , \ \hat{a} \ket{\psi_n} = \sqrt{\frac{\sigma^2}{2 \gamma}} \sqrt{n} \ket{\psi_{n-1}} 
\end{split}
\end{equation}
which closely resemble the properties of the analogous quantum mechanical operators \cite{griffiths2018}. 

\section{MSRJD path integral method}
\label{sec:MSRJD}

The MSRJD (Martin-Siggia-Rose-Janssen-De Dominicis) path integral description \cite{msr1973, janssen1976, dd1976, ddpeliti1978, hertz2016} of continuous stochastic systems described by SDEs like Eq. \ref{eq:bdaddnoise} offers an explicit formula for the transition probability $P(x, t; x_0, t_0)$ in terms of an infinite number of integrals. It resembles the phase space path integral \cite{shankar2011} from quantum mechanics, in that it involves integrating not just over all possible paths through state space, but also over auxiliary variables $p_j$. 

Two nice features of this approach that are worth highlighting are that (i) one can bypass the eigenfunction expansion and obtain the transition probability directly, and that (ii) no imagination (i.e. clever substitutions or tricks) is necessary. We \textit{just} need to calculate some integrals, and we will get our answer. For a derivation of this stochastic path integral, along with some additional discussion, see my earlier paper \cite{vastola2019}.  

The MSRJD path integral corresponding to Eq. \ref{eq:bdaddnoise} reads
\begin{equation} \label{eq:discreteMSRJDbd}
P = \lim_{N \to \infty} \int \frac{dp_N}{2\pi}\prod_{j = 1}^{N-1} \ \frac{dx_j dp_j}{2\pi} \ \exp{- \sum_{j = 1}^N \left[ i p_j \left( \frac{x_j - x_{j-1}}{\Delta t} - k + \gamma x_{j-1} \right) + \frac{1}{2} {p_j}^2 \sigma^2 \right] \Delta t} 
\end{equation}
where $\Delta t := (t - t_0)/N$, and where we are using $P$ as an abbreviation for $P(x_f, t_f; x_0, t_0)$. If we integrate out all of the momenta first, then we just have the Onsager-Machlup path integral, which is discussed in the next section; hence, we will try to integrate out the concentrations $x_j$ first. Change variables to $y_j = x_j - \mu$, so that Eq. \ref{eq:discreteMSRJDbd} becomes
\begin{equation} \label{eq:discreteMSRJDbdY}
P = \lim_{N \to \infty} \int \frac{dp_N}{2\pi}\prod_{j = 1}^{N-1} \ \frac{dy_j dp_j}{2\pi} \ \exp{- \sum_{j = 1}^N \left[ i p_j \left( \frac{y_j - y_{j-1}}{\Delta t} + \gamma y_{j-1} \right) + \frac{1}{2} {p_j}^2 \sigma^2 \right] \Delta t} 
\end{equation}
Define $C := 1 - \gamma \Delta t$. The action (i.e. the argument of the exponential) can be written
\begin{equation}
 - \sum_{j = 1}^N i p_j \left( y_j - C y_{j-1}  \right) + \frac{1}{2} {p_j}^2 \sigma^2 \Delta t = i C y_0 p_1 - i y_N p_N - i \sum_{j = 1}^{N-1} y_j \left( p_j - C p_{j+1}  \right) -  \frac{1}{2} \sum_{j=1}^N {p_j}^2 \sigma^2 \Delta t \ .
\end{equation}
We can easily integrate over $y_j$ for $j = 1, ..., N-1$ to obtain
\begin{equation}
 \int_{-\infty}^{\infty} \frac{dy_j}{2\pi} e^{- i y_j \left( p_j - C p_{j+1}  \right)} 
= \delta(p_j - C p_{j+1}) \ .
\end{equation}
Enforcing the $(N-1)$ delta function constraints leads to $p_1 = C p_2$, $p_2 = C p_3$, ..., $p_{N-1} = C p_N$. This means that we can write each $p_j$ in terms of the last one, $p_N$, via
\begin{equation} \label{eq:prelation}
p_j = C^{N-j} p_N \ .
\end{equation}
Using Eq. \ref{eq:prelation}, the remaining part of the action reads
\begin{equation}
\begin{split}
 i C y_0 p_1 - i y_N p_N -  \frac{1}{2} \sum_{j=1}^N {p_j}^2 \sigma^2 \Delta t = -i \left( y_N - C^N y_0 \right) p_N -  \frac{1}{2} \left[ \sigma^2 \Delta t \frac{1 - C^{2N}}{1 - C^2} \right] p_N^2 \ .
\end{split}
\end{equation}
All that remains of our path integral is an easily performed Gaussian integral:
\begin{equation}  \label{eq:MSRJDalmost}
\begin{split}
P =& \lim_{N \to \infty} \int \frac{dp_N}{2\pi} \ e^{-i \left( y_N - C^N y_0 \right) p_N -  \frac{1}{2} \left[ \sigma^2 \Delta t \frac{1 - C^{2N}}{1 - C^2} \right] p_N^2 } \\ =& \lim_{N \to \infty}  \sqrt{ \frac{1 - C^2}{2 \Delta t} \frac{1}{\pi \sigma^2 (1 - C^{2N}) }} \exp\left[ - \frac{\left( y_N - C^N y_0 \right)^2}{ \frac{2 \Delta t}{1 - C^2} \sigma^2  (1 - C^{2N})} \right]  \ .
\end{split}
\end{equation}
Using that $y_0 = x_0 - \mu$, $y_N = x - \mu$, and that
\begin{equation} \label{eq:goodlimits}
\begin{split}
& \lim_{N \to \infty} \frac{1 - C^2}{2 \Delta t} = \lim_{N \to \infty} \frac{1 - (1 - \gamma \Delta t)^2}{2 \Delta t} = \lim_{N \to \infty} \frac{2 \gamma \Delta t - \gamma^2 \Delta t^2}{2 \Delta t} = \gamma \\
& \lim_{N \to \infty} C^N = \lim_{N \to \infty} \left( 1 - \frac{T}{N}  \right)^N = e^{- \gamma T}  \ ,
\end{split}
\end{equation}
we can take the $N \to \infty$ limit of Eq. \ref{eq:MSRJDalmost} and recover Eq. \ref{eq:ptrans} as our final answer.

\section{Onsager-Machlup path integral method}

There is another path integral description of SDEs like Eq. \ref{eq:bdaddnoise} originally due to Onsager and Machlup \cite{OM1953pt1, OM1953pt2, graham1977, hertz2016}, which only involves integrals over state space. Surprisingly, despite it involving 	`fewer' integrals, explicit calculations are generally significantly harder. For a derivation of this path integral, see my earlier paper \cite{vastola2019}. 

We must compute
\begin{equation} \label{eq:OMbd}
P = \lim_{N \to \infty} \left( \frac{1}{\sqrt{2 \pi \sigma^2 \Delta t}} \right)^N \int \left[ \prod_{j = 1}^{N-1} dx_j \right] \exp\left\{ - \sum_{j = 1}^N \frac{\left[  \frac{x_j - x_{j-1}}{\Delta t} - k + \gamma x_{j-1} \right]^2}{2 \sigma^2} \Delta t   \right\}  
\end{equation}
where $\Delta t := (t - t_0)/N$. We will proceed by doing several changes of variables, hoping (eventually) to reduce the action to a simple form. We \textit{could} write one change of variables, but will do several, so it is clearer why we decided what we did. 

Define the constant $C := 1 - \gamma \Delta t$. Change variables from $x_j$ to $y_j$, then from $y_j$ to $z_j$, and then from $z_j$ to $w_j$ (for all $j = 0, 1, ..., N$), where
\begin{equation}
y_j := x_j - \mu \ , \ z_j := \frac{y_j}{\sqrt{\sigma^2 \Delta t}} \ , \ w_j := C^{-j} z_j
\end{equation}
so that the $j$th term in the action changes as
\begin{equation}
 \frac{\left[  \frac{x_j - x_{j-1}}{\Delta t} - k + \gamma x_{j-1} \right]^2}{2 \sigma^2/\Delta t}   \to \frac{\left[  y_j - y_{j-1} + \gamma \Delta t y_{j-1} \right]^2}{2 \sigma^2 \Delta t} \to \frac{\left[  z_j - C z_{j-1} \right]^2}{2} \to C^{2j} \left[  w_j - w_{j-1} \right]^2  \ .
\end{equation}
In terms of the $w_j$, Eq. \ref{eq:OMbd} reads
\begin{equation} \label{eq:OMbd_w}
P = \lim_{N \to \infty} \sqrt{\frac{C^{N(N-1)}}{(2\pi)^N \sigma^2 \Delta t} } \int \left[ \prod_{j = 1}^{N-1} dw_j \right] \exp\left\{ - \frac{1}{2} \sum_{j = 1}^N C^{2j} \left[  w_j - w_{j-1} \right]^2 \right\}  \ .
\end{equation}
Since the action can be written in the form
\begin{equation}
\begin{split}
 \sum_{j = 1}^N C^{2j} \left[  w_j - w_{j-1} \right]^2 
=& \left[ C^2 w_0^2 + C^{2N} w_N^2 \right]  + \left[ - 2 C^2 w_0 w_1 - 2 C^{2N} w_N w_{N-1} \right] \\
& + \sum_{j = 1}^{N-1} C^{2j} (1 + C^2) w_j^2 + \sum_{j = 1}^{N-2} - 2 C^{2(j+1)} w_j w_{j+1}
\end{split}
\end{equation}
we can write
\begin{equation}
 - \frac{1}{2} \sum_{j = 1}^N C^{2j} \left[  w_j - w_{j-1} \right]^2 = -\frac{1}{2} w^T A w + J \cdot w - \frac{\left[ C^2 w_0^2 + C^{2N} w_N^2 \right]}{2}
\end{equation}
where we define the matrix $A$ and the vector $J$ via
\begin{equation}
\begin{split}
A_{jj} &:= C^{2j} (1 + C^2) \ \text{ for } j = 1, ..., N-1 \\
A_{j+1,j} &:= - C^{2(j+1)} \ \text{ for } j = 1, ..., N-2 \\
A_{j,j+1} &:= - C^{2(j+1)} \ \text{ for } j = 1, ..., N-2 \\
A_{ij} &= 0 \ \text{ otherwise } \\
J &:= \begin{pmatrix} C^2 w_0 & 0 & \cdots & 0 & C^{2N} w_N \end{pmatrix}^T \ .
\end{split}
\end{equation}
At this point, we can invoke an integral often used in quantum field theory \cite{schwartz2014}:
\begin{equation} \label{eq:gaussianints}
\int \exp \left( - \frac{1}{2} x^T  A x + J \cdot x \right) d^nx = \sqrt{\frac{(2\pi)^n}{\det A}} \exp \left( \frac{1}{2} J^T A^{-1}  J \right)
\end{equation}
where $A$ is a real, symmetric, positive-definite matrix. Using Eq. \ref{eq:gaussianints}, we can write the transition probability as
\begin{equation} \label{eq:ptrans_matrix}
\begin{split}
P &= \lim_{N \to \infty} \sqrt{\frac{C^{N(N-1)}}{(2\pi)^N \sigma^2 \Delta t} } \sqrt{\frac{(2\pi)^{N-1}}{\det A}} \exp \left\{ \frac{1}{2} J^T A^{-1}  J - \frac{\left[ C^2 w_0^2 + C^{2N} w_N^2 \right]}{2} \right\} \\
&= \lim_{N \to \infty} \sqrt{ \frac{C^{N(N-1)}} {2\pi \sigma^2 \Delta t \det A} }\exp \left\{ \frac{1}{2} J^T A^{-1}  J - \frac{\left[ C^2 w_0^2 + C^{2N} w_N^2 \right]}{2} \right\} \ .
\end{split}
\end{equation}
In order to make sense of this, we need to compute two things: the determinant of $A$, and the quadratic form $J^T A^{-1}  J$. 

\subsection{Computing the determinant}

By writing out $A$ (an $N-1 \times N-1$ matrix) for various sizes, one can get some intuition by computing determinants
\begin{equation}
\begin{split}
\det A_1 &= C^2 (1 + C^2) \\
\det A_2 &= C^6 (1 + C^2 + C^4) \\
\det A_3 &= C^{12} (1 + C^2 + C^4 + C^6) \ .
\end{split}
\end{equation}
The determinant of $A_{j+1}$ can shown to be related to the determinants of $A_j$ and $A_{j-1}$ according to the recurrence relation
\begin{equation}
\det A_{j+1} = C^{2(j+1)} (1 + C^2) \det A_{j} - C^{4(j+1)} \det A_{j-1}
\end{equation}
which has solution
\begin{equation}
\det A_j = C^{j(j+1)} \left[ 1 + C^2 + \cdots + C^{2j} \right] \ .
\end{equation}
This can be proven by induction. Hence, the determinant of $A$ is
\begin{equation} \label{eq:detA}
\det A = C^{N(N-1)} \left[ 1 + C^2 + \cdots + C^{2(N-1)} \right] \ .
\end{equation}

\subsection{Computing the quadratic form}

Since only the first and last components of $J$ are nonzero, we have
\begin{equation} \label{eq:nonzeros}
J^T A^{-1} J = A^{-1}_{11} C^4 w_0^2 + A^{-1}_{N-1,N-1} C^{4N} w_N^2 + \left[ A^{-1}_{1,N-1} + A^{-1}_{N-1,1} \right] C^{2(N+1)} w_0 w_N \ .
\end{equation}
Because $A$ is symmetric, $A^{-1}_{1,N-1} = A^{-1}_{N-1,1}$. Components of $A^{-1}$ can be computed using the standard cofactor formula \cite{strang2016}
\begin{equation}
A^{-1}_{ij} = \frac{C_{ji}}{\det A}
\end{equation}
where $C_{ji}$ is the $(j, i)$ cofactor of $A$. Doing so, we obtain (after some lengthy calculations we do not record here)
\begin{equation} \label{eq:invcomps}
A^{-1}_{11} = \frac{1 - C^{2(N-1)}}{C^2 (1 - C^{2N})} \ , \ A^{-1}_{N-1,N-1} = \frac{1 - C^{2(N-1)}}{C^{2(N-1)} (1 - C^{2N})} \ , \ A^{-1}_{1,N-1} = \frac{1 - C^2}{C^2 (1 - C^{2N})}  \ .
\end{equation}
After substituting Eq. \ref{eq:invcomps} and Eq. \ref{eq:nonzeros} into Eq. \ref{eq:ptrans_matrix} and doing some algebra, we obtain
\begin{equation}
\begin{split}
& \frac{1}{2} J^T A^{-1} J - \frac{\left[ C^2 w_0^2 + C^{2N} w_N^2 \right]}{2}  = - \frac{C^{2N} (1 - C^2)}{2 (1 - C^{2N})} \left[ w_N - w_0\right]^2 \ .
\end{split}
\end{equation}
In terms of our original variable $x$, $w_N - w_0$ is
\begin{equation}
\begin{split}
w_N - w_0  = \frac{z_N - C^N z_0}{C^N}  = \frac{y_N - C^N y_0}{C^N \sqrt{\sigma^2 \Delta t}} = \frac{(x - \mu) - C^N (x_0 - \mu)}{C^N \sqrt{\sigma^2 \Delta t}}  
\end{split}
\end{equation}
so the argument of the exponential in Eq. \ref{eq:ptrans_matrix} reads
\begin{equation}
\begin{split}
& \frac{1}{2} J^T A^{-1} J - \frac{\left[ C^2 w_0^2 + C^{2N} w_N^2 \right]}{2} = - \frac{1 - C^2}{2 \Delta t} \frac{1}{ \sigma^2 (1 - C^{2N})} \left[ (x - \mu) - C^N (x_0 - \mu) \right]^2  \ .
\end{split}
\end{equation}

\subsection{Finishing the calculation}

Using Eq. \ref{eq:detA}, the prefactor in Eq. \ref{eq:ptrans_matrix} can be written 
\begin{equation}
\begin{split}
 \sqrt{ \frac{C^{N(N-1)}} {2\pi \sigma^2 \Delta t \det A} } = \sqrt{ \frac{1 - C^2} {2\pi \sigma^2 \Delta t  \left[ 1 - C^{2N} \right]} }  
\end{split}
\end{equation}
so that all we have left to do is compute
\begin{equation} \label{eq:ptrans_rematrix}
\begin{split}
P &= \lim_{N \to \infty} \sqrt{ \frac{1 - C^2} {2\pi \sigma^2 \Delta t  \left[ 1 - C^{2N} \right]} }   \exp \left\{ - \frac{1 - C^2}{2 \Delta t} \frac{1}{ \sigma^2 (1 - C^{2N})} \left[ (x - \mu) - C^N (x_0 - \mu) \right]^2  \right\} \ .
\end{split}
\end{equation}
Reusing the limits in Eq. \ref{eq:goodlimits} from the MSRJD path integral section, we again derive Eq. \ref{eq:ptrans}.

\section{WKB/semiclassical method}
\label{sec:WKB}

The Wentzel–Kramers–Brillouin (WKB) or semiclassical approach \cite{schulman1981, gutzwiller1971, girard1992, littlejohn1992} to approximating the quantum mechanical propagator can straightforwardly be adapted to approximate the transition probability $P(x, t; x_0, t_0)$ by applying the usual arguments to the Onsager-Machlup path integral. For one view of semiclassical approximations in stochastic systems, see Assaf and Meerson \cite{assaf2017}, although we will proceed somewhat differently. Note that this approach is somewhat distinct from WKB-type approaches intended to estimate $P_{ss}(x)$ only \cite{risken1996, zhouli2016}. 

Almost as in quantum mechanics, we have
\begin{equation} \label{eq:WKB}
P(x, t; x_0, t_0) \approx N(t) \exp\left[ - S_{cl} \right] 
\end{equation}
where $N(t)$ is a time-dependent prefactor, and where the `classical' action $S_{cl}$ is defined as
\begin{equation}
S_{cl}(x, x_0, T) := \int_{t_0}^t L \ dt' = \int_{t_0}^t \frac{(\dot{x} - k + \gamma x)^2}{2 \sigma^2} \ dt'
\end{equation}
i.e. as the time integral of the Onsager-Machlup Lagrangian (c.f. Eq. \ref{eq:OMbd}) along the most likely trajectory $x(t)$. The Lagrangian corresponding to the most likely transition path is
\begin{equation}
L = \frac{(\dot{x} - k + \gamma x)^2}{2 \sigma^2} 
\end{equation}
and its Euler-Lagrange equation \cite{goldstein2001, elsgolc2012} can be shown to reduce to
\begin{equation}
\ddot{x} - \gamma^2 x = - k \gamma \ .
\end{equation}
This has the general solution
\begin{equation}
x(t) = c_1 e^{\gamma (t-t_0)} + c_2 e^{- \gamma (t-t_0)} + R
\end{equation}
where $c_1$ and $c_2$ are arbitrary constants and $R$ is the particular solution. The particular solution can be found by substitution to be $\mu$, while $c_1$ and $c_2$ can be found to be (after enforcing $x(t_0) = x_0$ and $x(t) = x$)
\begin{equation}
\begin{split}
c_1 &= \frac{x - (x_0 - \mu) e^{- \gamma T} - \mu}{e^{\gamma T} - e^{- \gamma T}} \\
c_2 &= x_0 - \mu - c_1 \ .
\end{split}
\end{equation}
Hence, the classical action $S_{cl}$ is
\begin{equation}
S_{cl} = \int_{t_0}^t L \ dt' = \frac{1}{2 \sigma^2} \int_{t_0}^t (\dot{x} - k + \gamma x)^2 \ dt' \ .
\end{equation}
Note that
\begin{equation}
\dot{x} - k + \gamma x = \gamma c_1 e^{\gamma (t-t_0)} - \gamma c_2 e^{- \gamma (t-t_0)}  - k + \gamma c_1 e^{\gamma (t-t_0)} + \gamma c_2 e^{- \gamma (t-t_0)} + k = 2 \gamma c_1 e^{\gamma (t-t_0)}
\end{equation}
so that we have
\begin{equation}
S_{cl} = \frac{2 \gamma^2 c_1^2}{\sigma^2} \int_{t_0}^t e^{2 \gamma (t' - t_0)} \ dt' = \frac{2 \gamma^2 c_1^2}{\sigma^2} \frac{\left[ e^{2 \gamma T} - 1 \right]}{2 \gamma} = \frac{\gamma}{\sigma^2} \frac{[x - x_0 e^{- \gamma T} - \mu (1 - e^{- \gamma T})]^2}{1 - e^{- 2 \gamma T}} \ .
\end{equation}
At this point, we would normally have to calculate a functional determinant in order to evaluate $N(t)$ \cite{kleinert2009}; however, we can take a shortcut and just guess that $N(t)$ is what we would get from naively normalizing the transition probability. Hence, the semiclassical estimate for the transition probability is
\begin{equation}
P(x, t; x_0, t_0) = \sqrt{\frac{\gamma}{\pi \sigma^2 (1 - e^{- 2 \gamma T})}} \exp\left[ - S_{cl} \right] 
\end{equation}
which, amazingly, exactly matches Eq. \ref{eq:ptrans}. In general, it will only be an approximation.

\section{Approximating the CME solution}
\label{sec:CMEcomp}

The CME for the chemical birth-death process (Eq. \ref{eq:CME}) can be solved by the Jahnke and Huisinga ansatz \cite{jahnke2007}, by the Doi-Peliti path integral approach \cite{peliti1985}, or (in principle) by a different path integral description of the CME \cite{vastolaCLE2019}. It can also be solved using arguments similar to the ones we used in Sec. \ref{sec:eigexpansion} and \ref{sec:ladder}, which yield the eigenfunction expansion solution
\begin{equation}
P^{CME}(x, t; x_0, t_0) =  \sum_{n=0}^\infty \frac{\mu^n}{n!} C_n(x_0, \mu) C_n(x, \mu) P_{ss}^{CME}(x) e^{- \gamma n T} \ ,
\end{equation}
where the $C_n$ are Charlier polynomials, which satisfy \cite{nilsson2012, koekoek2010}
\begin{equation}
\sum_{n = 0}^{\infty} C_n(x, a) C_m(x, a) = \delta_{n m} \frac{n! e^{\mu}}{\mu^n}
\end{equation}
\begin{equation} \label{eq:charlierlimit}
\lim_{a \to \infty} (2 a)^{n/2} C_n(a + z \sqrt{2a}, a) = (-1)^n H_n(z) \ .
\end{equation}
We described in Sec. \ref{sec:intro} how the CME (Eq. \ref{eq:CME}) is related to our SDE (Eq. \ref{eq:bdaddnoise}); in this section, we will show that the \textit{solution} to the former reduces to the solution of the latter in the large $\mu$ limit. 

First, we will verify that the steady state solution to the CME reduces to Eq. \ref{eq:pss} in the large $\mu$ limit. The steady state probability distribution corresponding to Eq. \ref{eq:CME} is
\begin{equation}
P^{CME}_{ss}(x) = \frac{\mu^x e^{-\mu}}{x!} \ .
\end{equation}
Assuming that $x$ is large and applying Stirling's approximation, we have
\begin{equation}
\begin{split}
P^{CME}_{ss}(x) &\approx \frac{1}{\sqrt{2 \pi x}}\frac{\mu^x e^{-\mu}}{\left( \frac{x}{e} \right)^x} = \frac{1}{\sqrt{2 \pi x}} \exp\left\{ - x \log\left( \frac{x}{\mu} \right) + (x - \mu)  \right\} \ .
\end{split}
\end{equation}
Assume that deviations from the mean are relatively small, i.e. that $(x - \mu)/\mu \ll 1$. Then
\begin{equation}
\begin{split}
\log\left( \frac{x}{\mu} \right) &= \log\left( 1 + \frac{x - \mu}{\mu} \right) \approx \frac{x - \mu}{\mu} - \frac{(x - \mu)^2}{2 \mu^2} \\
\end{split}
\end{equation}
Now we have
\begin{equation}
\begin{split}
P^{CME}_{ss}(x) &\approx  \frac{\exp\left\{ - x \left[ \frac{x - \mu}{\mu} - \frac{(x - \mu)^2}{2 \mu^2} \right] + (x - \mu)  \right\}}{\sqrt{2 \pi x}} =  \frac{\exp\left\{ - \frac{(x - \mu)^2}{\mu} - \frac{x(x - \mu)^2}{2 \mu^2} \right\}}{\sqrt{2 \pi x}} \ .
\end{split}
\end{equation}
At this point we approximate $x$ as $\mu$ in two places: the prefactor, and the second term inside the exponential. In the first place, writing $\sqrt{2 \pi x} \approx \sqrt{2 \pi \mu}$ is harmless, because the $x$-dependence of the function is dominated by the exponential anyway. In the second place, writing $x/\mu \approx 1$ avoids having a term third order in $x$, and second order in $\mu$ (both of which are inappropriate for our crude approximation). We get
\begin{equation}
\begin{split}
P^{CME}_{ss}(x) &\approx \frac{1}{\sqrt{2 \pi \mu}} \exp\left\{ - \frac{(x - \mu)^2}{\mu} - \frac{\mu(x - \mu)^2}{2 \mu^2} \right\} = \frac{1}{\sqrt{2 \pi \mu}} \exp\left\{ - \frac{(x - \mu)^2}{2 \mu} \right\} 
\end{split}
\end{equation}
which is just what we derived from the Fokker-Planck equation (c.f. Eq. \ref{eq:pss}) with $\sigma^2 = 2 k$ (see Sec. \ref{sec:intro}). 

Finally, using Eq. \ref{eq:charlierlimit}, we can show that
\begin{equation}
\lim_{\mu \to \infty} (2 \mu)^{n/2} C_n(x, \mu) = (-1)^n H_n \left(\frac{x - \mu}{\sqrt{2 \mu}} \right)
\end{equation}
and hence that 
\begin{equation}
P^{CME}(x, t; x_0, t_0) \approx \sum_{n=0}^\infty \frac{1}{2^n n!} H_n \left(\frac{x_0 - \mu}{\sqrt{2 \mu}} \right) H_n \left(\frac{x - \mu}{\sqrt{2 \mu}} \right)  P_{ss}(x) e^{- \gamma n T} \ ,
\end{equation}
in the large $\mu$ limit (c.f. Eq. \ref{eq:Hexpansion}). Summing the propagator as in Sec. \ref{sec:propagatorsum}, we recover Eq. \ref{eq:ptrans} with $\sigma^2 = 2 k$. 

\section{Conclusion}

We solved the chemical birth-death process with additive noise analytically in eight qualitatively different ways. In doing so, we implicitly examined the strengths and weaknesses of many different approaches to analytically solving problems in stochastic dynamics. The intuition we get from the previous calculations is that these methods fall into four broad categories: (i) simple and has potential to generalize (the MSRJD path integral, method of characteristics, CME solution approximation); (ii) simple but hard to generalize (direct SDE solution, semiclassical approximation); (iii) lengthy but much easier to generalize (the eigenfunction expansion methods); and (iv) hard and hard to generalize (the Onsager-Machlup path integral). In particular, we imagine that the approaches we just identified as generalizable can readily be applied to systems consisting of many interacting Ornstein-Uhlenbeck-like processes (i.e. the additive noise SDE equivalent of a system of monomolecular reactions, whose CME is known to be solvable analytically \cite{jahnke2007}).

\section{Acknowledgments}

This work was supported by NSF Grant \# DMS 1562078.

\bibliography{addnoise_bib, re_pathint}

\end{document}